# An Algorithm for Finding Convex Hulls of Planar Point Sets

Gang Mei, John C.Tipper
Institut für Geowissenschaften – Geologie
Albert-Ludwigs-Universität Freiburg
Freiburg im Breisgau, Germany
{gang.mei, john.tipper}@geologie.uni-freiburg.de

Nengxiong Xu
School of Engineering and Technology
China University of Geosciences (Beijing)
Beijing, China
xunengxiong@yahoo.com.cn

*Abstract*—This paper presents an alternate choice of computing the convex hulls (CHs) for planar point sets. We firstly discard the interior points and then sort the remaining vertices by x- / y- coordinates separately, and later create a group of quadrilaterals (e-Quads) recursively according to the sequences of the sorted lists of points. Finally, the desired CH is built based on a simple polygon derived from all e-Quads. Besides the preprocessing for original planar point sets, this algorithm has another mechanism of discarding interior point when form e-Quads and assemble the simple polygon. Compared with three popular CH algorithms, the proposed algorithm can generate CHs faster than the three but has a penalty in space cost.

*Keywords—Convex hull; point set; extreme points*

## I. INTRODUCTION

Finding the convex hull (CH) of point sets is a fundamental issue in computational geometry, computer graphics, robotics, etc. Some of the most popular algorithms of building CHs include Graham scan [1], Jarvis march [2], Monotone chain [3], Quickhull [4], Divide–and–Conquer [5] and Incremental [6].

In [7], the algorithms for producing CHs were classified into two categories: *graph traversal* and *incremental* ones. The graph traversal algorithms firstly find some vertices of CH and then intend to identify the remaining points and edges by traversing it in some approaches. Representative ones are Graham scan [1], Jarvis march [2] and Monotone chain [3].

Incremental algorithms conduct CHs by finding an initial CH and then inserting or merging the remaining points, edges or even sub CHs into current CH sequentially or recursively to obtain the final CHs. Quickhull [4], Divide–and–Conquer [5] and Incremental [6] can be deemed as some of this class.

Recently, several novel algorithms are developed to obtain CH for point set: Franěk and Matoušek[9] present a polynomial-time algorithm for the D-convex hull of a finite point set in the plane. In [10], new properties of CH are derived and then used to eliminate concave points to reduce the computational cost.

Clarkson, Mulzer and Seshadhri [11] describe an algorithm for computing planar convex hulls in the self-improving model. A combinatorial structure, hypermaps, is used to model planar subdivisions of the plane for designing a functional algorithm which computes the convex hull of a finite set of points incrementally [12].

Liu and Wang [13] propose a reliable and effective CH algorithm based on a technique named Principle Component Analysis for preprocessing the planar point set. A fast CH algorithm with maximum inscribed circle affine transformation is developed in [14]; and another two quite fast algorithms are both designed based on GPU [15, 16].

In this paper, we present an alternate algorithm to produce the convex hull for points in the plane. The main ideas of the proposed algorithms are as follows. A planar point set is firstly preprocessed by discarding its interior points; secondly the remaining points are sorted by x- and y- coordinates separately and divided into several sub-regions. And later we create a group of quadrilaterals (e-Quads) recursively according to the sequences of the sorted lists of points. Finally, the desired CH is built based on a simple polygon derived from all e-Quads.

## II. THE PROPOSED ALGORITHM

### A. Overview of the Algoirhtm

Our algorithm can be divided into five steps:

Step 1: Find 4 extreme points with minimum or maximum coordinates of x or y, and then discard the interior points located inside the polygon consisted of the above 4 points.

This simple procedure of preprocessing is widely used to efficiently filter the input points and then effectively reduce the computation cost. An improved version of this method which attempts to find 8 extreme points is adopted in [10].

Step 2: Sort the remaining points by x- and y- coordinates separately, and divide the points to being in 4 sub-regions.

The extreme points with minimum or maximum x, and minimum or maximum y of a point set are denoted as $X_{min}$, $X_{max}$, $Y_{min}$ and $Y_{max}$, respectively. The quadrilateral formed by above four extreme points is defined as e-Quad.

Step 3: Form all e-Quads recursively according to the sequences of the sorted lists of points.

Firstly, we create the first e-Quad according to the sorted lists of points, and then remove the four vertices of the e-Quad from the current point set; after that, next group of four extreme points of the rest points can also be found to form the second e-Quad. This procedure composed of finding, creating, and removing will be repeated recursively until no points left.

This research was supported by the Natural Science Foundation of China (Grant Numbers 40602037 and 40872183) and Fundamental Research Funds for the Central Universities of China.

In this step, when find the extreme points according to the sorted lists of points, some candidate extreme points may locate inside the last formed e-Quad and these points must be discarded directly.

Step 4: Assemble a simple polygon based on all e-Quads.

An edge chain can be formed in each sub-region by adding and updating proper vertices of each e-Quad according to the sequence of being created. The final simple polygon can be assembled by connecting the individual edge chains in different sub-regions.

Step 5: Calculate the CH of the simple polygon created in Step 4 by Melkman's CH algorithm [8] to be the CH of the input point set.

*B. Discarding and Dividing*

The preprocessing for the input point set by discarding the interior points is quite efficient and effective. Normally, the extreme points are selected to form a polygon and then all points are indentified whether locate inside the polygon.

In [10], eight extreme points are selected to create the polygon for testing interior points. In this paper, we use the simple and commonly-used version: only four extreme points, the leftmost, the bottommost, the rightmost and the topmost, are selected to form the original polygon for indentifying.

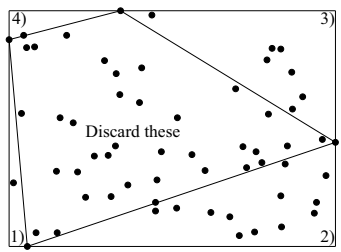

Figure 1. Discarding and dividing

As shown in Fig.1, a rectangular region can be defined by the first group of extreme points. This region can also be deemed as the smallest Axis-Aligned Bounding Box (AABB) of the set of points.

Without considering the polygonal area formed by the extreme points, the rest part of the AABB can be divided into four sub-regions. This division of the AABB plays no role in this step but will be quite important to assemble the simple polygon in Step 4.

*C. Forming all e-Quads*

The first e-Quad is obviously the polygon formed in Step 1 for detecting interior points. After creating this e-Quad, its four vertices will be removed from the current sorted lists of points. Similarly, next group of four extreme points of the rest points can also be found to form the second e-Quad.

In this step, when find the extreme points according to the sorted lists of points, the candidate extreme points in this moment have to be checked whether they are inside the last created e-Quad. If the points locate inside the last formed e-Quad, they must be also discarded.

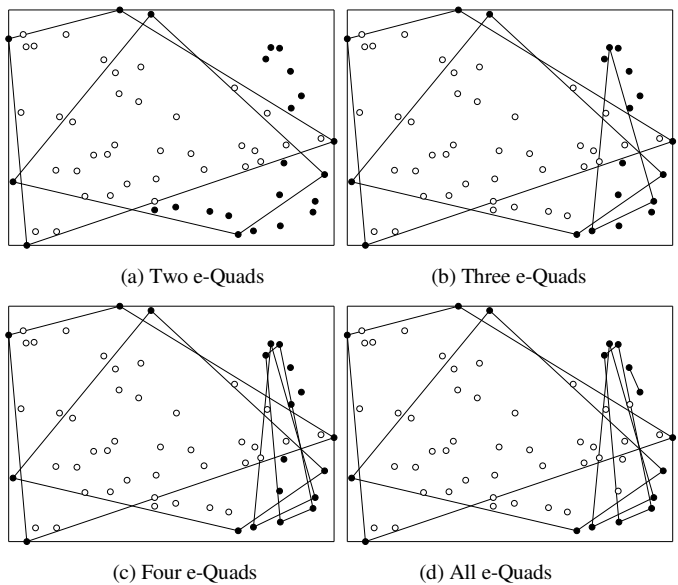

(a) Two e-Quads  (b) Three e-Quads

(c) Four e-Quads  (d) All e-Quads

Figure 2. Creating all e-Quads for a planar point set

This procedure composed of finding valid (not in the last e-Quad) extreme points, creating a new e-Quad, and removing vertices of the new e-Quad from current sorted lists of points will be repeated recursively until no points left (Fig. 2).

Noticeably, not all interior points inside the current e-Quad can be detected and then discarded. The detection of whether a point locates inside the last created e-Quad terminates when the desired four extreme points have been found.

*D. Asssembing the Simple Polygon*

After creating all e-Quads, the edge chains in four sub-regions can be assembled separately by adding and updating proper vertices of each e-Quad according to the sequence of being created, as shown in Fig. 3.

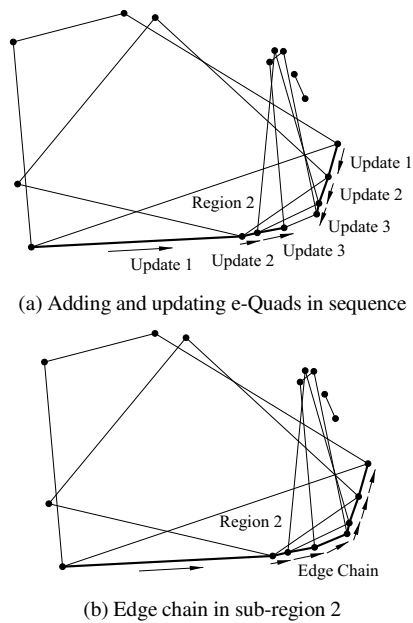

(a) Adding and updating e-Quads in sequence

(b) Edge chain in sub-region 2

Figure 3. Creating edge chains in different sub-region 2

**Remark 1** Only part of the four vertices of an e-Quad need to be indentified and then merged to form the edge chains.

In Step 1, we divide the AABB of a point set into four sub-regions. The sub-region 1 is defined by the leftmost and the bottommost points. When create edge chain in sub-region 1, only the vertices $X_{min}$ and $Y_{min}$ of each e-Quad need to be checked whether they are in this sub-region and then adopted to form the edge chain when inside.

Assuming an e-Quad located inside the sub-region 1, as shown in Fig. 4a, the 3$^{rd}$ and 4$^{th}$ vertex of an e-Quad, $X_{max}$ and $Y_{max}$ obviously locate in the left side of the edge chain formed by the leftmost, $X_{min}$, $Y_{min}$ and the bottommost. Hence, $X_{max}$ and $Y_{max}$ should not be the vertices of the final CH and thus can be discarded directly.

When the vertices, $X_{max}$ and $Y_{max}$ are not in the sub-region 1, as shown in Fig. 4b, it's unnecessary to consider them in sub-region 1 but to indentify them in the regions 3 and 4, respectively.

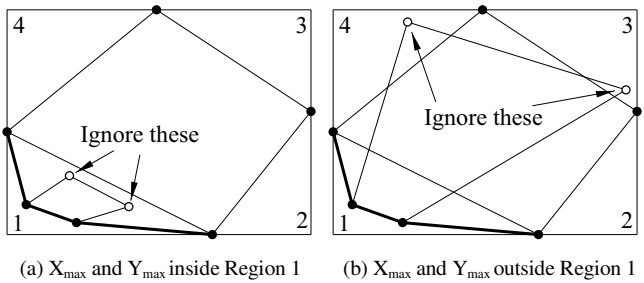

(a) $X_{max}$ and $Y_{max}$ inside Region 1  (b) $X_{max}$ and $Y_{max}$ outside Region 1

Figure 4.   Selecting proper points of each e-Quad when create the edge chain in sub-region 1

The mechanism of selecting proper vertices of each e-Quad can be extended from the sub-region 1 to other three regions (listed in Table 1).

TABLE I.   SELECTION OF PROPER VERTICES IN DIFFERENT SUB-REGIONS

| Sub-region | Defined by | Proper vertices of each e-Quad |
|---|---|---|
| 1 | Leftmost, bottommost | $X_{min}$ and $Y_{min}$ |
| 2 | Bottommost, rightmost | $Y_{min}$ and $X_{max}$ |
| 3 | Rightmost, topmost | $X_{max}$ and $Y_{max}$ |
| 4 | Topmost, leftmost | $Y_{max}$ and $X_{min}$ |

**Remark 2** A point used twice in an e-Quad (degenerate cases) can only be adopted once to form the edge chains.

In degenerate cases, some points are duplicated as part of the vertices of an e-Quad. If these points are chosen to form an edge chain, the edge chain will self-intersect, and can not be used to be a part of the final simple polygon.

Noticeably, when select the proper vertices of each e-Quad to form the edge chains in the sub-regions, every vertex is accepted once in the best-cases while twice in the worst-case. Obviously, we can conclude that a vertex of each e-Quad is used at most twice; therefore, the time complexity of forming all edge chains is $O(n)$.

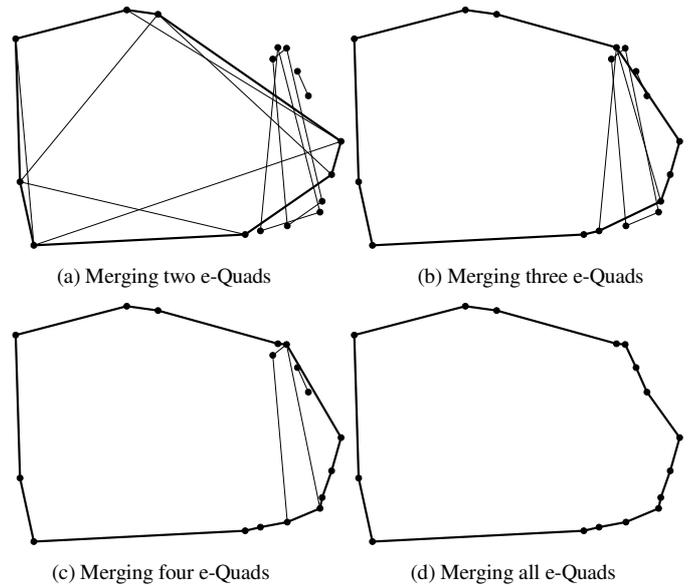

(a) Merging two e-Quads  (b) Merging three e-Quads

(c) Merging four e-Quads  (d) Merging all e-Quads

Figure 5.   Forming simple polygon

The final simple polygon can be assembled by dynamically connecting the individual edge chains in different sub-regions, as shown in Fig. 5. Normally, four edge chains can be formed in sub-regions. When there are only 3 or 2 sub-regions in the degenerate cases, edge chains are still created in each sub-region as displayed in Fig. 3, and then also be assembled to obtain the desired simple polygon.

*E. Calculating the CH*

After obtaining the simple polygon, we prefer to adopt the Melkman's algorithm [8] to find the CH of the simple polygon. The CH shown in Fig. 6 is also the desired one for the original planar point set.

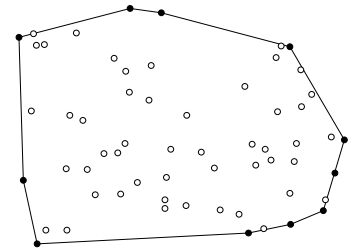

Figure 6.   Convex hull of a set of planar points

Melkman's CH algorithm computes the convex hull of a simple polygonal chain (or a simple polygon) in linear time. This effective algorithm does not need any preprocessing, and just processes vertices sequentially once. It only uses a double-ended queue (also called deque) to store an incremental hull for the vertices already processed.

III.   PERFORMANCES

In this section, we first analyze the complexity of our algorithm, and then create several groups of sample points to test and compare the speed with popular algorithms including Graham scan [1], Jarvis march [2] and Monotone chain [3].

## A. Complexity Analysis

In Step 1, the preprocessing of discarding interior points needs $O(n)$; and the sorting in Step 2 runs in $O(n\log n)$; since each points is used at most twice when form all e-Quads and assemble simple polygon, the time complexity of Steps 3 and 4 is $O(n)$; finally, calculating the CH by Melkman's algorithm costs $O(n)$ time. Hence, overall time complexity is $O(n\log n)$.

The streams for inputting point sets and outputting the CH cost $O(n)$ space; and the intermediate variables for storing the sorted lists of points and the simple polygon also need $O(n)$ space. Thus, the space complexity of this algorithm is $O(n)$.

## B. Experimental Tests

The original data points are created randomly in a rectangle or a cycle. For each size of points set, 25 samples are created and then tested with the mentioned four algorithms. The final running-time is the average value of 25 computational costs.

The running-time of finding the CHs for different size of point sets uniformly distributed in a rectangle (Table 2) or a cycle (Table 3) is compared separately.

We have implemented our algorithm and compared with three popular algorithms under VC++2010 on the platform of Windows 7 using the same machine which features an Intel(R) Core(TM) i5 M450 2.40GHz processor and 2 GB of memory.

TABLE II. RUNNING-TIME FOR PLANAR POINTS UNIFORMLY LOCATED IN RECTANGULAR REGION (UNIT: SECOND)

| Size of Points | Running-time of CH algorithms (/s) | | | |
|---|---|---|---|---|
| | *Monotone Chain* | *Jarvis March* | *Graham Scan* | *Our Algorithm* |
| 50,000 | 0.264 | 0.291 | 0.109 | 0.130 |
| 100,000 | 0.540 | 0.476 | 0.226 | 0.264 |
| 200,000 | 1.126 | 1.017 | 0.471 | 0.514 |
| 500,000 | 3.011 | 2.161 | 1.316 | 1.626 |
| 1,000,000 | 5.928 | 3.782 | 2.582 | 3.060 |

TABLE III. RUNNING-TIME FOR PLANAR POINTS UNIFORMLY LOCATED IN CIRCULAR REGION (UNIT: SECOND)

| Size of Points | Running-time of CH algorithms (/s) | | | |
|---|---|---|---|---|
| | *Monotone Chain* | *Jarvis March* | *Graham Scan* | *Our Algorithm* |
| 50,000 | 0.256 | 1.316 | 0.106 | 0.104 |
| 100,000 | 0.533 | 3.305 | 0.225 | 0.216 |
| 200,000 | 1.085 | 8.039 | 0.472 | 0.457 |
| 500,000 | 2.840 | 25.834 | 1.273 | 1.233 |
| 1,000,000 | 5.903 | 60.591 | 2.747 | 2.689 |

## IV. CONCLUSTION AND DISCUSSION

We present an alternate algorithm for finding the CHs of planar point sets. We firstly discard the interior points and then sort the rest vertices to form e-Quads, and finally calculate the CH based on the simple polygon derived from all e-Quads.

The algorithm can in further discard interior points in the procedure of forming e-Quads, and ignore some kind of vertices of each e-Quad when assemble the simple polygon. This mechanism of rejecting points can help improve speed.

Compared with three popular CH algorithms, for the point sets uniformly distributed in a rectangle, the algorithm is faster than Monotone chain and Jarvis march but slower than Graham scan; for the point sets uniformly distributed in a cycle, the proposed algorithm is faster than the above three.

The shortcoming of this algorithm is that the space cost is expensive due to that several intermediate variables need to be allocated for sorting and storing the sorted sequence.

ACKNOWLEDGMENT

The corresponding author Gang Mei would like to thank Chun Liu at Universität Kassel for sharing several interesting and valuable ideas.